# Polarization based modulation of splitting ratio in femtosecond laser direct written directional couplers

Zhi-Kai Pong[*], Bangshan Sun, Patrick S. Salter, and Martin J. Booth

*Department of Engineering Science, University of Oxford, Parks Road, Oxford OX1 3PJ, UK*
*\*zhikai.pong@eng.ox.ac.uk*

**Abstract:** This work characterizes a phenomenon in direct laser written directional couplers where the splitting ratio for output light is dependent on the input polarization state. In general, for laser written waveguides, different coupling strengths exist for different polarization states of the input light. If the linear polarization state of the input light is not aligned with one of the symmetry axes of the system, an additional amplitude beating is imposed on the transfer of light in directional couplers of different interaction length. We present results for in-plane and out of plane directional couplers, which are supported by theoretical analysis. These new results provide insights for understanding and controlling polarization properties of directional couplers and larger photonic circuits.

## 1. Introduction

Polarization is a crucial property of light for encoding information for applications in photonic computing and communication [1]. It is therefore important to understand how optical devices respond to different polarizations in order to have full control on manipulating different states and preserving the quality of encoded information.

Previous work on femtosecond laser direct written (FLDW) photonic circuits has demonstrated different approaches to utilize polarization-dependent properties of integrated optical devices to let them act in lieu of bulk components, such as polarization beam splitters [2], waveplates [3] and retarders [4]. However, the characterization typically only considers vertical and horizontal linear polarization inputs, but not for arbitrary linear polarizations. Moreover, the devices were mostly coplanar, which gives an incomplete description of device behavior when considering three-dimensional structures. Szameit et al. investigated how out of plane geometries affect coupling constants, but did not take input polarization into account [5]. Sansoni et al. fabricated out of plane directional couplers specifically so that the coupling was the same for horizontal and vertically polarized light [6]. Nevertheless, it was focused on finding a specific operating point for a fixed interaction region length.

We present a general polarization-dependent effect where the orientation of linearly polarized incident light dictates the splitting ratio for each output port of a directional coupler and can modulate the maximum value attainable. To the best of our knowledge, this phenomenon has not been characterized in literature before. Moreover, we further demonstrate the effect in directional couplers extended to three-dimensional structures, where the two arms are not at the same depth from the substrate surface. The experimental results are in agreement with the theoretical description of the coupling.

The FLDW method [7] focuses an ultrafast pulse laser to a highly confined volume, modifying the material through nonlinear absorption processes. In glass, this results in an increase in local refractive index [8]. When tracing the laser focus through the glass block, the resultant modified region therefore acts as a waveguide for light transmission. One of the strengths of FLDW inside a suitable material lies in its ability to create three-dimensional structures integrated in a single device. Recent work has been undertaken to utilize this capability to demonstrate topological and quantum phenomena [9-11]. However, to fully exploit this feature, it is crucial to understand how these out-of-plane geometries change the properties of the devices and the transmission for different polarization states.

Ultrafast laser-written directional couplers are investigated in this research as they are fundamental building blocks of optical circuits [12]. A waveguide directional coupler splits incident light into different output ports depending on the interaction length of the coupling region. It is formed by bringing two waveguides close to each other. Their function relies on evanescent coupling, which occurs when two waveguides are in close proximity and their evanescent fields overlap [13]. In this paper, the waveguide arm coupled to the input light will be referred to as the primary waveguide, and the other arm referred to as the secondary waveguide. The splitting ratio is the proportion of light intensities in the primary and secondary waveguide outputs with respect to the input light of a directional coupler. The ratio varies with respect to the interaction region length of a directional coupler.

Fig. 1(a) shows the orientation of the input polarization angle θ with respect to the y-direction (vertical) of the directional couplers. Note that the red dashed line indicates the interaction region for evanescent coupling of the two arms of a directional coupler. Fig. 1(b) is a schematic diagram of the x-z plane view of a directional coupler with input light at one end of the primary waveguide and outputs at the other ends of the two arms. $L_0$ is the interaction region length for evanescent coupling and $d$ is the separation distance between the two arms. Fig. 1(c) and Fig. 1(d) shows the waveguides being offset at an angle which we call ϕ, which is used when studying the effect on structures that are not coplanar.

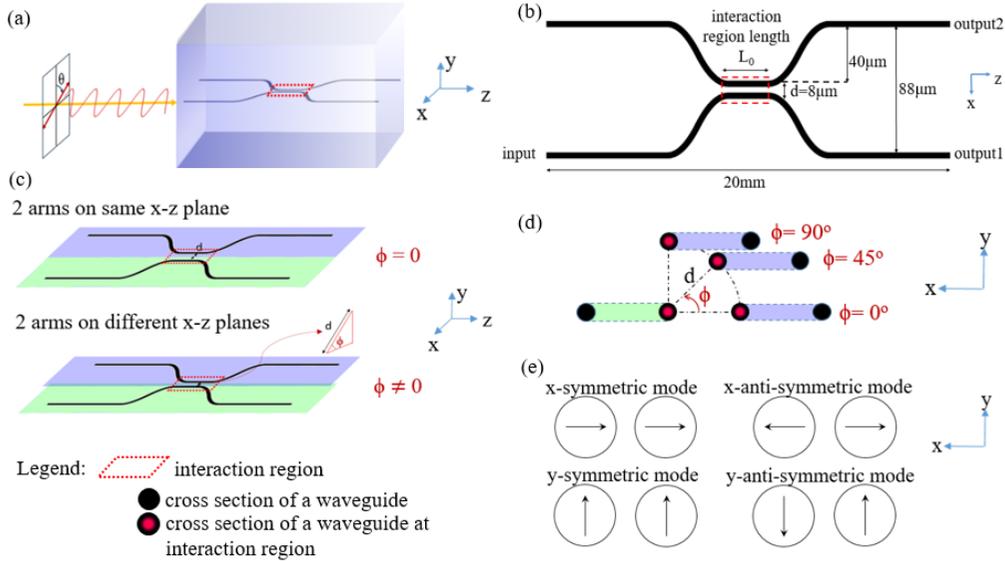

Fig. 1. (a) Schematic showing the input light polarization angle θ with respect to the vertical y-direction of the directional couplers. (b) 2D top view of a schematic directional coupler with two waveguides on same x-z plane. (c) 3D view of two waveguides lying on same or different x-z planes with a constant separation distance $d$ slightly offset at an angle ϕ at the interaction region – the upper diagram shows when ϕ = 0 and the lower when ϕ ≠ 0. (d) 2D cross-section view on x-y plane of two waveguides at different ϕ at the interaction region. (e) The four fundamental modes of parallel waveguides at the interaction region. Arrows in the same direction indicate the fields in both waveguides are in phase, and opposite directions indicate out of phase. The arrows point along the polarization axes [14].

We characterized the polarization effects on splitting ratio as a function of the interaction region length. This allows us to understand how the performance of directional coupler devices depends on polarization. The theoretical study for the characterization of the dependence of the polarization orientation angle and relative position of the waveguide arms predicts an additional modulation effect. This gives insights into how we can control the fabrication in order to improve the performance of devices by better understanding the polarization effects.

Below, Section 2 describes the background and theoretical study. Section 3 describes details of the experimental setup and the directional coupler fabrication process. Section 4 reports and discusses the experimental results and the results of extension to 3D structures. Section 5 gives the concluding remarks.

## 2. Background and theory

The splitting ratio $r$ indicates the proportions of light intensities in the primary and secondary waveguide outputs with respect to the input light of a directional coupler. According to coupled-mode theory [15, 16], the expression for $r$ is given by

$$r = \frac{P_1}{P_1 + P_2} = \sigma^2 \sin^2\left(\frac{C}{\sigma}L_0\right) \qquad (1)$$

where $P_1$ and $P_2$ are the output power intensities of the primary and secondary waveguides respectively, $C$ is the coupling coefficient between the two waveguides, $\sigma$ is a dephasing term that depends on the waveguide asymmetry and $L_0$ is the length of the interaction region. The dephasing term $\sigma$ is given by

$$\sigma = \frac{1}{\sqrt{1 + \left(\frac{\Delta\beta}{2C}\right)^2}} \qquad (2)$$

where $\Delta\beta = |\beta_1 - \beta_2|$ is the difference of the propagation constants of the two waveguides [17]. Therefore, if there is some general asymmetry in the waveguide refractive index distribution, the maximum power coupling ratio will be reduced to a fraction of the original. Nevertheless, the expression predicts a sinusoidal behavior with respect to the interaction region length. However, Eq. (1) does not fully consider polarization effects and asymmetry for different input polarization states.

When the input polarization is aligned with the principal axes of symmetry of the cross-section geometry (along x- and y-axis for ϕ = 0° coplanar directional couplers as illustrated in Fig. 1(a)), the light propagates through the arms with a specific mode, which can be viewed as a superposition of symmetric and anti-symmetric modes (see Fig. 1(e)). Power exchange between the primary and secondary arms can then be explained in terms of the beating between the symmetric and anti-symmetric modes. However, when the input polarization is not aligned along either of these principal axes, both x-polarized and y-polarized modes are present, and the situation is more complicated. The x/y symmetric/anti-symmetric modes all have different propagation constants, therefore these four modes will all contribute differently to the total propagating wave.

Assume the two waveguides lie in the xz-plane, where z is the direction of light propagation through the waveguides (Fig. 1(a)). Birefringence caused by non-ideal device fabrication by FLDW, which often produces waveguides with non-circular cross-sections [18], leads to non-identical effective refractive indices of the primary and secondary waveguides for different input polarization states. Furthermore, the geometric configuration of the two waveguides leads to further asymmetry, particularly in the strain field, such that the propagation constants of the symmetric mode and the anti-symmetric mode of the x- and y-polarized light are therefore also different [19]. This difference leads to a change in the splitting ratio when the input polarization changes and is inherent to the geometry of the directional couplers.

We also simulated the effect in COMSOL Multiphysics software. A 3D model of the structure was created in the software, with the geometry of the directional coupler as two parallel cylindrical waveguides next to each other. The effective refractive indices of the four modes closest to that of the waveguide core were computed, which were the symmetric/anti-symmetric modes of the x/y polarizations. The simulation results showed that the effective refractive indices for these four modes are different, which supports the hypothesis that these modes will propagate through the structure differently hence affecting the coupling between the two waveguides.

Theoretically, the input light can be expressed as a sum of the symmetric and antisymmetric modes with equal amplitude, and the beating between the modes creates a modulated envelope which manifests as the power exchange between the two arms of the directional coupler. The rate of power exchange is slightly different for polarized input light along the two principal axes, hence there will be an additional modulation to the splitting ratio for other polarizations. For an arbitrary linearly polarized light with the angle $\theta$ between the polarization axis and the vertical y-direction (Fig. 1(a)), the modulation is most pronounced at $\theta = \pi/4$. The derivation of this result is shown below.

## 2.1 Derivation for linearly x-polarized and y-polarized input light

Let the propagation constants of the x-symmetric (even) mode be denoted as $k_{xe}$ and that of the x-anti-symmetric (odd) mode be denoted as $k_{xo}$. Similarly, the propagation constants for symmetric mode and anti-symmetric mode of the y-polarized modes are $k_{ye}$ and $k_{yo}$ respectively. Let $E_{xe}(z)$ represent the electric field amplitude for the propagating electric field corresponding to the x-symmetric mode as a function of $z$, omitting the time-varying component which is assumed to be the same for all terms. Similarly, for the other modes

$$E_{xe}(z) = \cos(k_{xe}z), E_{xo}(z) = \cos(k_{xo}z),$$
$$E_{ye}(z) = \cos(k_{ye}z), E_{yo}(z) = \cos(k_{yo}z) \tag{3}$$

where $z$ is along the propagation direction.

Consider input light polarized along the y-direction coupled to the primary waveguide arm. This input light can be expressed as a sum of the symmetric and antisymmetric modes with equal amplitude. For unit amplitude input, at $z = 0$ we have $P_1 = 1$ and $P_2 = 0$. Assuming losses by absorption and scattering are negligible, we have $P_1 + P_2 = 1$ for all $z$ since power is conserved. For simplicity, in the following, the power expressions $P_x$, $P_y$ and $P_\theta$ refer to the output power intensities of the x-polarized, y-polarized and linearly polarized input light with polarization axis at an angle $\theta$ to the vertical y-axis. They represent power in the primary waveguide arm i.e $P_1$ in Eq. (1). Note that $P_2$ in the secondary waveguide arm is therefore simply $1 - P_1$, being consistent with the notation in Eq. (1).

Let the wave amplitude in the primary waveguide corresponding to the y-polarized input light be $E_y$. Then

$$\begin{aligned} E_y(z) &= \frac{1}{2}\big(E_{ye}(z) + E_{yo}(z)\big) \\ &= \frac{1}{2}\big[\cos(k_{ye}z) + \cos(k_{yo}z)\big] \\ &= \cos\left(\frac{k_{ye} + k_{yo}}{2}z\right)\cos\left(\frac{k_{ye} - k_{yo}}{2}z\right) \end{aligned} \tag{4}$$

As $k_{ye}$ and $k_{yo}$ have similar but unequal values, the second cosine term is of much lower frequency than the first one. The expression therefore represents a modulated sinusoid with an envelope corresponding to the second term. The power flow in the waveguide is obtained by integrating the wave amplitude, therefore it is proportional to the square of the envelope. The power intensity $P_y$ corresponding to the y-polarized input with unit amplitude is given by:

$$P_y(z) = \cos^2\left(\frac{k_{ye} - k_{yo}}{2} z\right) \tag{5}$$

Similarly,

$$E_x(z) = \cos\left(\frac{k_{xe} + k_{xo}}{2} z\right) \cos\left(\frac{k_{xe} - k_{xo}}{2} z\right) \tag{6}$$

and

$$P_x(z) = \cos^2\left(\frac{k_{xe} - k_{xo}}{2} z\right) \tag{7}$$

The envelope shows how the power flow in the primary arm changes with distance. We define quantities $\kappa_x$ and $\kappa_y$ as follows:

$$\kappa_x = k_{xe} - k_{xo}$$
$$\kappa_y = k_{ye} - k_{yo} \tag{8}$$

They can be interpreted as the propagation constants of the wave envelopes which manifests as the power exchange between the two arms of the directional coupler.

### 2.2 Derivation for arbitrary linearly polarized input light

Now consider an arbitrary linearly polarized light with polarization axis along θ coupled to the primary waveguide arm. Interference effects between modes (e.g. between $k_{xe}$ and $k_{ye}$) are insignificant compared to cross-talk in directional couplers [20]. The overall intensity can therefore be found by decomposing the polarization vector into x- and y- components. Since the wave vector corresponding to this input is a vector sum of $E_x$ and $E_y$ components, for input light polarized along θ with unit amplitude we have:

$$P_\theta(z = L_0) = P_x(z = L_0) \cos^2 \theta + P_y(z = L_0) \sin^2 \theta \tag{9}$$

where the output power intensity is measured at the output of a directional coupler with interaction distance $L_0$.

Since the effective propagation constants for the envelopes of $E_x$ and $E_y$ are different, there will also be beating between the envelopes of the two waves. To demonstrate this behavior, we will give an example for θ = π/4 where the beating is most pronounced when $E_x$ and $E_y$ have equal amplitudes:

$$\begin{aligned} P_{\frac{\pi}{4}}(z = L_0) &= P_x \cos^2\left(\frac{\pi}{4}\right) + P_y \sin^2\left(\frac{\pi}{4}\right) \\ &= \frac{1}{2}(P_x + P_y) \\ &= \frac{1}{2}\left[\cos^2\left(\frac{\kappa_x}{2} L_0\right) + \cos^2\left(\frac{\kappa_y}{2} L_0\right)\right] \\ &= \frac{1}{2}\left[\frac{1 + \cos(\kappa_x L_0)}{2} + \frac{1 + \cos(\kappa_y L_0)}{2}\right] \\ &= \frac{1}{2}\left[1 + \cos\left(\frac{\kappa_x + \kappa_y}{2} L_0\right) \cos\left(\frac{\kappa_x - \kappa_y}{2} L_0\right)\right] \end{aligned} \tag{10}$$

The second cosine term shows an additional modulation to the sinusoidal variation of the splitting ratio with increasing interaction region length. This variation has a period between that of pure x- and y- polarized input and the modulation envelope has a wavenumber proportional to $\frac{\kappa_x - \kappa_y}{2}$. This additional modulation is important to consider when using integrated photonic circuits comprising directional couplers with light of variable input polarization state. It should be noted that there is also a change of the polarization state for the output light from the mixing between different polarization states. This effect arises due to the birefringence in the isolated waveguides, and could be compensated by a single birefringent element as long as the birefringence is identical in the two arms [21].

## 3. Experimental Setup

### 3.1 Fabrication System

The laser used in waveguide writing was a frequency-doubled regeneratively amplified Yb:KGW laser (Light Conversion Pharos SP-06-1000-pp) at 1MHz repetition rate, 168 fs pulse duration, and 515 nm wavelength. A combination of a motorized rotating half-wave plate and a polarization beam splitter was used to adjust the average laser power.

A liquid-crystal on silicon spatial light modulator (SLM) (Hamamatsu Photonics X10468-09 (X)) was used to correct for system and sample aberrations [22]. The SLM is imaged by a 4-f system onto the pupil plane of the objective (0.5 NA; 20×; Zeiss Plan Neofluar), which then focused the laser into the specimen placed on the sample stage.

The devices were fabricated in borosilicate glass (Corning EAGLE 2000). The glass sample was fixed on a three-axis air bearing translation stage (AerotechABL10100L, xy-motion; ANT95-3-V, z-motion) for moving the sample relative to the laser focus.

### 3.2 Fabrication parameters

Unless otherwise specified, the waveguides were written with a single scan at 150μm depth from the surface, 8mm/s scanning speed, and 110nJ pulse energy measured at the objective pupil.

After completion of the waveguide fabrication, the samples were polished by using a sequence of 30μm, 9μm, 3μm and 1μm polishing films. A layer of at least 150μm glass was polished off both the input and output facets of the chip.

A directional coupler was created fabricating two waveguides separated by 88μm at the input and output facets, then brought together in the middle of the sample with 8μm separation in the interaction region (Fig. 1(b)).

Multiple sets of directional couplers were fabricated with different interaction region lengths from 2mm to 10mm in 0.5mm steps to investigate variation of splitting ratio between two arms when polarization properties of input light were changed.

### 3.3 Characterization setup

To characterize and measure the performance of the fabricated devices, a testing rig was used to couple light from a fiber into the waveguides. A fiber-coupled laser source (Thorlabs S1FC780PM) was coupled to a polarization-maintaining single mode fiber (Thorlabs PM630-HP (PANDA)) and the light emitted has a wavelength of 785nm. The other end of the fiber was held by a fiber rotator (Thorlabs HFR007) which was mounted on a 6-axis stage (Thorlabs MAX600/M Series).

The glass sample was mounted onto a 3-axis translation stage (Newport M-562-XYZ-LH). The other end of the waveguides was imaged onto a CCD camera (Baumer TXD-14) using an objective (Olympus ULWD MS Plan 80x/0.75NA) and an achromatic lens (Thorlabs AC254-

100-A-ML). The fiber was butt-coupled into the input of waveguide sample, with the fiber-tip in proximity (<1μm) to minimize coupling losses.

For directional coupler characterization, the fiber was coupled to one of the input waveguide arms. The splitting ratio of each directional coupler was then measured by integrating over each output using pixel values obtained from the images captured with the CCD camera. Fig. 2(a) shows the output facet of the primary and secondary waveguides under LED illumination. Fig. 2(b) shows a sample image of the light output at the ends captured with the CCD camera.

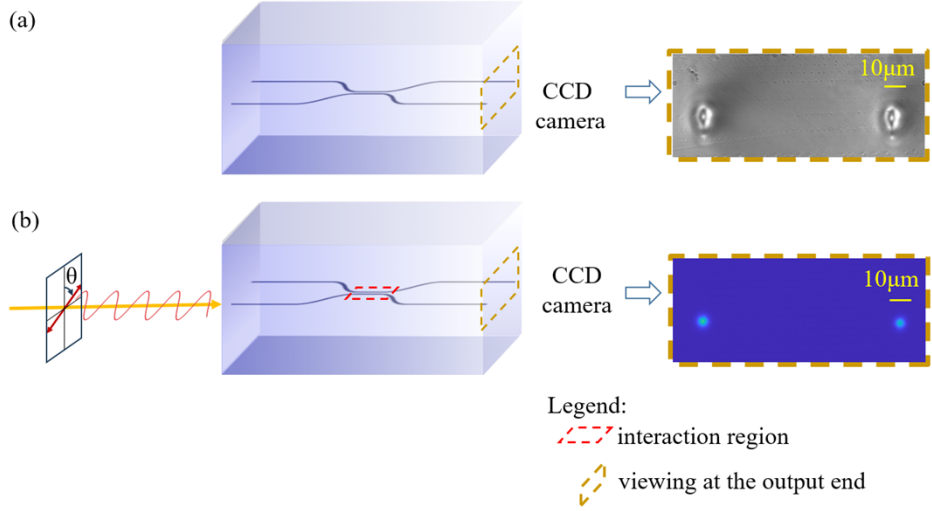

Fig. 2. View of experiment measurements at the output end of a directional coupler: (a) output facet of primary and secondary waveguides showing the cross-section of fabricated waveguides which are non-circular and elongated along the y-direction. (b) Microscopic image of laser light guiding modes at the two outputs.

## 4. Experimental results

We obtained measurements of the variation of splitting ratio against interaction region length when the input light had a linear input polarization at different angles θ to the vertical y-axis. We found that the coupling variation had a maximum period when the input polarization θ was at 0° and a minimum at 90°. The difference in period indicates a difference in the coupling coefficient $C$, which is in agreement with the theoretical analysis presented in Section 2. This observed polarization dependence is common to all directional coupler configurations provided there is a difference between $\kappa_x$ and $\kappa_y$.

From the derivation in Section 2, we have shown that we only need $\kappa_x$ and $\kappa_y$ to know the variation for all other values of θ. The values for $\kappa_x$ and $\kappa_y$ are thus inferred from the experimental data measured at θ = 0° and 90° by fitting the measured data to a sinusoid using Eq. (1) and minimizing least-square errors, with an additional phase that takes into account the coupling in the bend region that connects the input and output to the interaction region [23]. Let the period of the fitted variation be $T$, then $\kappa_x$ and $\kappa_y$ are given by

$$\kappa_x = \frac{2\pi}{T_x}$$
$$\kappa_y = \frac{2\pi}{T_y}$$
(11)

We present in the following figures with the experimental data points and corresponding fitting or simulation curves. The main source of measurement error arises from the coupling between the fiber and the waveguide input. Measurements were perfomed on the set of directional couplers for at least 5 times, and the measured data points are shown with error bars which indicate the range of measured values.

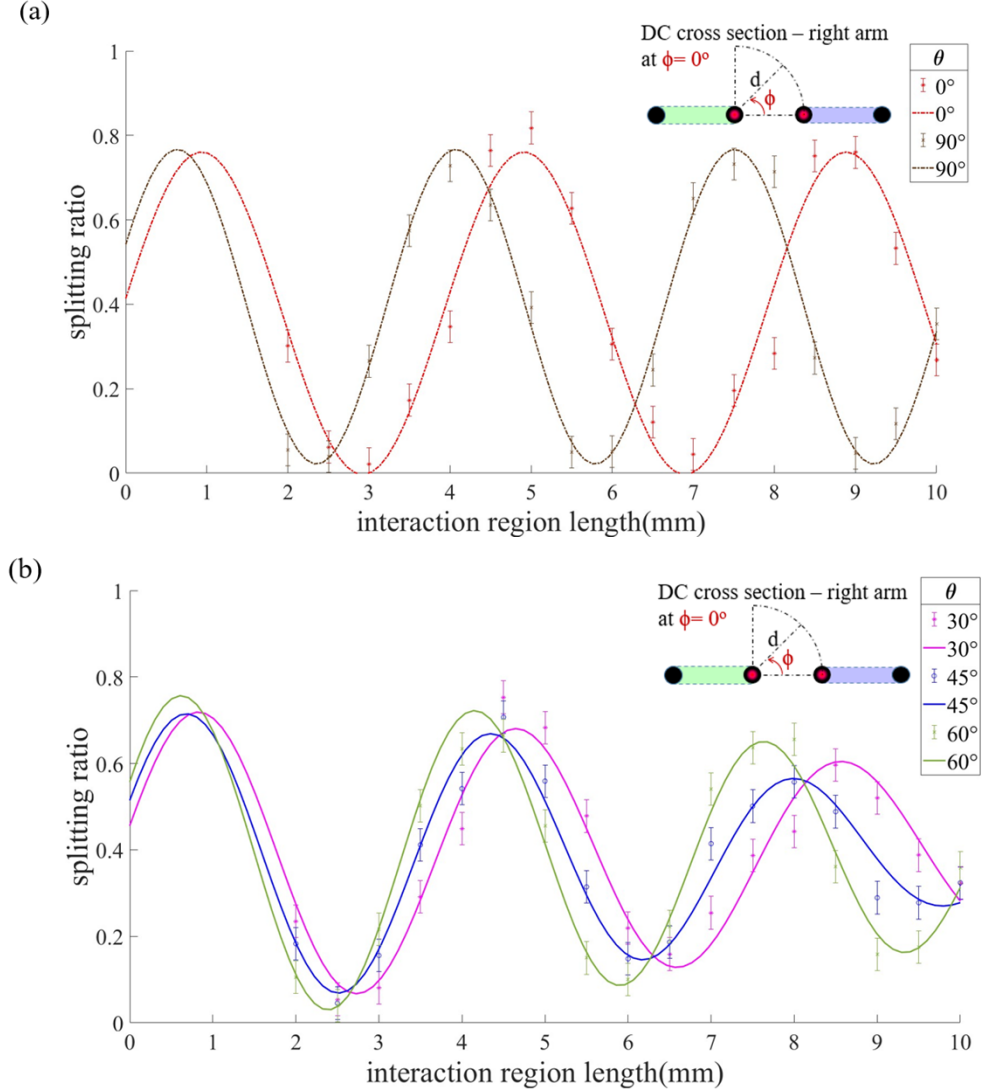

Fig. 3. Splitting ratio against interaction region length of laser-written directional couplers (DC) with input polarization axis angle θ to vertical, and ϕ at 0°. (a) Measured splitting ratios and least-squares fitted sinusoidal curves for θ at 0° and 90°. (b) Measured splitting ratios and theoretical predicted curves for θ at 30°, 45° and 60°. Error bars indicate measurement errors.

Fig. 3(a) show the fitted sinusoidal curves for θ = 0° and θ = 90°, which are used to infer the values of $\kappa_x$ and $\kappa_y$ using Eq. (11). The values are found to be $\kappa_x = 1.83 mm^{-1}$ and $\kappa_y = 1.58 mm^{-1}$, which are typical of similar devices in literature [24, 25]. Fig. 3(b) shows the predicted curves for θ = 30°, 45° and 60° using Eq. (9). The results show that the curves predicted by theory are in good agreement with the measured data. They are no longer purely sinusoidal. A trend of decreasing amplitudes over several cycles is observed, and the decrease is most rapid at 45°.

The sum of least-squares error for sinusoidal curve fitting in Fig. 3(a) is 0.070, whereas the errors for the theoretical predictions in Fig. 3(b) are 0.038, 0.029 and 0.020 respectively for θ = 30°, 45° and 60°. This shows that the prediction matches the measured result with high accuracy. The higher error in curve-fitting is likely due to error in the polarization angle measurement since the fiber rotator (Thorlabs HFR007) has scale marks with 5° increments and was adjusted manually, hence slight modulation effects due to the angle difference were not taken into account with a pure sinusoid.

To further verify our understanding, we also fabricated waveguide arms of the DC that were slightly offset such that they have the same separate distance $d$ but are on different xz-planes and at an angle ϕ from each other on the cross-section view (Fig. 1(c) and (d)). By compensating for the depth-dependent aberrations [26], we ensure that the cross-sections of the waveguides remain uniform at different depths. Comparing results for in-plane and out-of-plane directional couplers enables further understanding of the dependence on polarization.

Fig. 4 is similar to Fig. 3 except that instead of ϕ = 0°, the right arm was at ϕ = 90° from the left arm with the same separation distance $d$. Similar to Fig. 3(b), Fig. 4(b) shows a trend of decreasing amplitudes over several cycles, and the decrease is most rapid at 45°. This confirms that the observed polarization phenomenon is generalizable to out-of-plane geometrical configurations of directional couplers.

We also note that when ϕ = 90°, the splitting ratio has a limited upper bound of around 0.5. From Fig. 4(a), the values of κ are found to be $\kappa_x = 3.73 mm^{-1}$ and $\kappa_y = 3.56 mm^{-1}$. These values are approximately doubled from Fig. 3. This is due to the asymmetry of the waveguides causing dephasing between the coupling of the two waveguides. This effect is reflected by the σ term in Eq. (1). As a result, the amplitude of splitting ratio is reduced by a factor of $\sigma^2$, and the period of the variation is also reduced by a factor of σ. The value of σ can be found by finding the ratio between κ for ϕ = 0° and ϕ = 90°. This gives $\sigma_x = 0.50$ and $\sigma_y = 0.44$. From Fig. 2(a), we note that the modified refractive index region of the fabricated waveguide cross-section was elongated along the y-direction. Since the separation of the two waveguides from center to center is fixed at 8μm, the modified refractive index region of the upper waveguide overlaps partially with the lower one. This changes the refractive index profile in the interaction region between the two waveguides and also affects the mode shape, hence the coupling between the two arms is no longer symmetric.

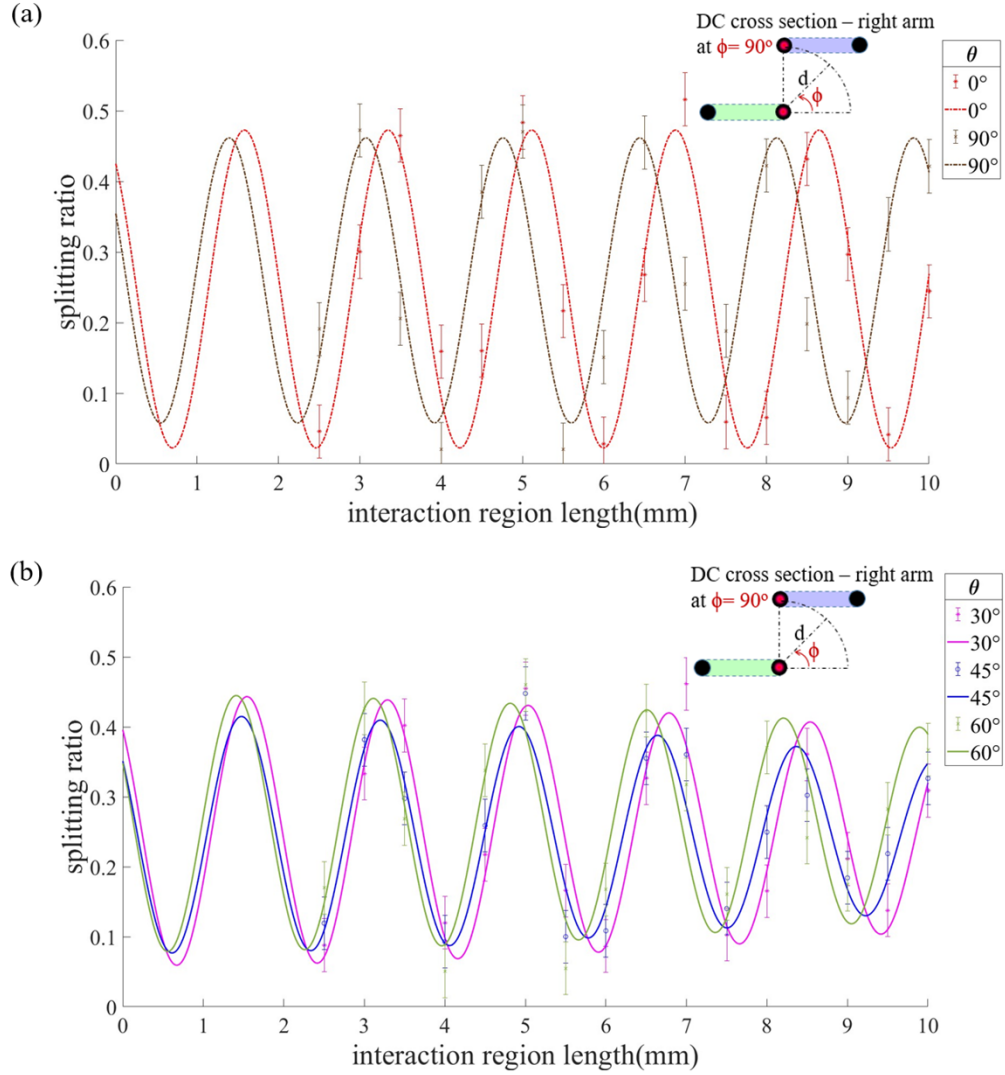

Fig. 4. Splitting ratio against interaction region length of laser-written DC with input polarization axis angle θ to vertical, and ϕ at 90°. (a) Measured splitting ratios and least-squares fitted sinusoidal curves for θ at 0° and 90°. (b) Measured splitting ratios and theoretical predicted curves for θ at 30°, 45° and 60°. Error bars indicate measurement errors.

Fig. 5 shows the case for ϕ = 45°, where the geometry of the interaction region is effectively rotated by 45°. We observe that the unmodulated polarizations are now at θ = 45° and 135° along the principal axes. The values of $\kappa$ are found to be $\kappa_{45°} = 1.80 mm^{-1}$ and $\kappa_{135°} = 1.77 mm^{-1}$. The amplitude of the splitting ratio for angles not along the principal axes displays a modulation similar to the case of ϕ = 0°.

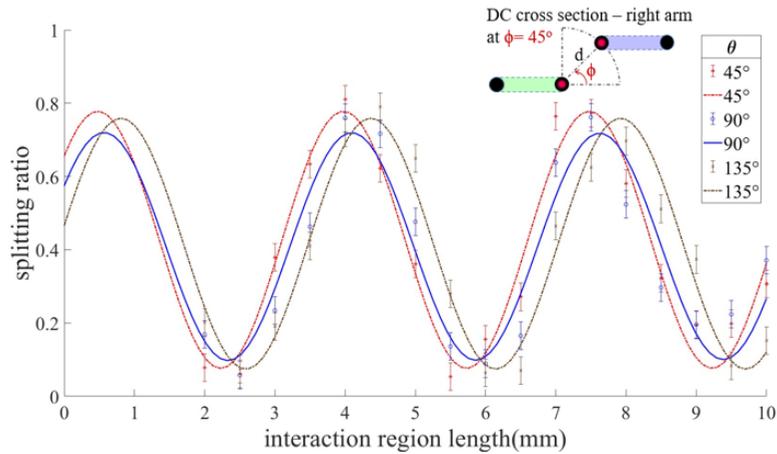

Fig. 5. Splitting ratio against interaction region length of laser-written DC with input polarization axis angle θ to vertical, and ϕ at 45°: dashed curves - least-squares fitted sinusoidal curves for θ at 45° and 135°; solid curve - theoretical predicted curves for θ at 90°. Error bars indicate measurement errors.

Given the experimental results as shown above, we were able to correctly predict the variation for arbitrary linear input polarization given the geometry of the directional couplers. A simple prediction procedure can be performed as follows: Firstly we measure the $\kappa_x$ and $\kappa_y$ for any directional coupler geometry from finding the period of variation for polarization along the principal axes by minimizing least-square errors of data points for fitted sinusoidal curves. We can then use Eq. (9) to infer the variation for all other angles. This enables much quicker characterization and verification of directional coupler devices for more efficient prototyping and design processes.

## 5. Conclusion

In summary, we have analyzed theoretically and observed in experiments that there is a modulation of the splitting ratio when the input polarization does not align with the principal axes of the geometrical configuration. We also showed how different geometrical configurations of relative angular offset affect this variation. Supported by experimental results, we demonstrated that this modulation phenomenon is generalizable to out-of-plane geometrical configurations. We believe that this understanding is an important contribution to further optimizations of fabrication parameters in 3D photonic circuits. Combined with ways of controlling the birefringence of individual waveguides [27-29], this knowledge potentially enables new designs of polarization-related applications that fully utilizes the three-dimensional capabilities of ultrafast direct laser writing.

**Disclosures**

The authors declare no conflicts of interest.

**Data availability**

Data underlying the results presented in this paper are not publicly available at this time but will be released through the Oxford University Research Archive.